\documentclass{aa}
\usepackage{graphicx}
\usepackage{txfonts}

\newcommand{\beq}{\begin{equation}}
\newcommand{\beqa}{\begin{eqnarray}}
\newcommand{\eeq}{\end{equation}}
\newcommand{\eeqa}{\end{eqnarray}}
\newcommand{\etal}{{\it et al. }}

\newcommand{\lsim}{\la}
\newcommand{\gsim}{\ga}
\newcommand{\mL}{\mathrm{L}}
\newcommand{\mS}{\mathrm{S}}
\newcommand{\mLS}{\mathrm{LS}}
\newcommand{\mo}{\mathrm{o}}
\newcommand{\mi}{\mathrm{i}}
\newcommand{\mE}{\mathrm{E}}
\newcommand{\ms}{\mathrm{s}}
\newcommand{\me}{\mathrm{e}}
\newcommand{\mn}{\mathrm{n}}
\newcommand{\mk}{\mathrm{k}}

\begin{document}
   \authorrunning{R. Takahashi, T. Suyama and S. Michikoshi}
   \titlerunning{Scattering of Gravitational Waves by Lens Objects}

   \title{Scattering of Gravitational Waves by the Weak Gravitational
           Fields of Lens Objects}

   \subtitle{}

   \author{R. Takahashi$^1$, T. Suyama$^2$ and S. Michikoshi$^2$}

   \offprints{R. Takahashi}

   \institute{$^1$ Division of Theoretical Astronomy, National
               Astronomical Observatory of Japan, Mitaka, 
               Tokyo 181-8588, Japan \\
              $^2$ Department of Physics, Kyoto University,
                Kyoto 606-8502, Japan 
             }

   \date{Received; accepted}

   \abstract{
     We consider the scattering of gravitational waves by the
   weak gravitational fields of lens objects. 
   We obtain the scattered gravitational waveform
   by treating the gravitational potential of the lens
   to first order, i.e. using the Born approximation.
   We find that the effect of scattering on the
   waveform is roughly given by the Schwarzschild radius of the lens
   divided by the wavelength of gravitational wave
   for a compact lens object.
   If the lenses are smoothly distributed,
   the effect of scattering is of the order of the convergence field
   $\kappa$ along the line of sight to the source. 
   In the short wavelength limit, the amplitude is magnified by
   $1+\kappa$, which is consistent with the result in weak gravitational
   lensing.

   }

   \maketitle
%

\section{Introduction}

Ground-based laser interferometric detectors of gravitational waves
 such as LIGO, VIRGO, TAMA and GEO are currently in operation to search
 for astrophysical sources such as neutron star
 binaries, black hole binaries and supernovae
 (e.g. Cutler \& Thorne \cite{ct02}).
The gravitational wave signals from these binaries are extracted from the
 data using matched filtering with a gravitational waveform template.
If the gravitational waves pass near massive compact
 objects or pass through intervening inhomogeneous mass distribution,
 the gravitational waveform is changed due to the scattering (or the
 gravitational lensing) by the gravitational potential of these objects.
The gravitational waves do not directly interact with
 matter (e.g. Thorne \cite{t87}), but the gravitational lensing occurs
 in the same way as it does for electromagnetic waves. 
In this letter, we investigate the effects of scattering by lens objects
 on the gravitational waveform.

In the gravitational lensing of light, the scattering is
 discussed in terms of gravitational lensing under the geometrical optics
 approximation, which is valid because the wavelength is much smaller
 than the typical size of lens objects. 
But in the case of gravitational waves, since the wavelength is much
 larger than that of light, geometrical optics is not valid
 in some cases.
If the wavelength is larger than the Schwarzschild radius of the
 lens, wave optics should be used (Peters \cite{p74}; Ohanian \cite{o74};
 Bontz \& Haugan \cite{bh81}; Thorne \cite{t83};
 Deguchi \& Watson \cite{dw86}).
This condition is rewritten as $M \lsim 10^3 M_\odot
 (f/10^2 {\mbox{Hz}})^{-1}$,
 where $10^2$Hz is the typical frequency of gravitational waves for
 ground-based detectors. 
Hence, we use wave optics in this letter.

In the geometrical optics for the lensing of light, the strong and weak
 lensing are distinguished by the convergence field $\kappa$ which is the
 ratio of surface density of lens $\Sigma$ to a critical density
 $\Sigma_\mathrm{cr} \sim c^2/G D_\mathrm{L}$, where $D_\mathrm{L}$ is
 the distance to the lens
 (Kaiser \cite{k92}; Bartelmann \& Schneider \cite{bs01}).
In the strong lensing regime, $\kappa \gsim 1$, the multiple images of distant
 source are formed.
But the strong lensing probability is small, $\sim 0.1 \%$, for high
 redshift sources.
Hence, the weak lensing approximation, $\kappa \ll 1$, is valid for
 most sources.
We use the weak field approximation in the wave optics.

In the past, Peters (\cite{p74}) studied the scattering by a point mass lens
 and a thin sheet of matter in the weak field approximation,
 and obtained the scattered waveform for these lens models.
The gravitationally lensed waveform ( which is the solution of wave
 equation (\ref{weq}) ) was given in Schneider, Ehlers \& Falco
 (\cite{sef92}), Sec.4.7 and 7,
 using the diffraction integral under the thin lens approximation.
Recently, several authors have been studying wave optics in the
 gravitational
 lensing of gravitational waves using this integral (Nakamura \cite{n98};
 Nakamura \& Deguchi \cite{nd99}; Ruffa \cite{r99}; Takahashi \& Nakamura
 \cite{tn03}; Yamamoto \cite{y03}; Macquart \cite{m04}; Takahashi
 \cite{t04} ).
In this letter, we present another method to derive the solution of
 Eq.(\ref{weq}) in the weak-field limit.
We treat the gravitational field of lens to first order, i.e. using
 the Born approximation, and discuss its validity.
We use units of $c=G=1$. 

\vspace*{0.9cm}


\section{Gravitational Waves Propagating through Weak Gravitational Fields}

We consider the scattering of gravitational waves by the weak
 gravitational fields of lens objects.
We consider the scattering of scalar waves, instead of gravitational
 waves, since the basic equation is the same as the scalar field wave
 equation (see Sect. IIC of Peters \cite{p74}). 
The gravitational fields of the lens objects are described by the metric
 $g_{\mu \nu}$ as $ds^2 = g_{\mu \nu} dx^{\mu} dx^{\nu} = -\left( 1+2U 
 \right) dt^2 + \left( 1-2U \right) d\mathbf{r}^2$, where $U(\mathbf{r})$
 is the gravitational potential of the lens.
Since the propagation equation of the scalar wave is
 $\partial_\mu \left( \sqrt{-g} g^{\mu \nu} \partial_\nu \phi \right)=0$,
 we have the wave equation as
\beq
  \left( \nabla^2 + \omega^2 \right) \tilde{\phi}(\mathbf{r})
  = 4 \omega^2 U(\mathbf{r}) \tilde{\phi}(\mathbf{r}),
\label{weq}
\eeq  
where $\omega$ is the frequency of the gravitational waves and
 we set $\phi(t,\mathbf{r}) = \tilde{\phi}(\mathbf{r}) \me^{\mi \omega t}$.

\begin{figure}
  \hspace{1.8cm}
  \includegraphics[width=5.cm]{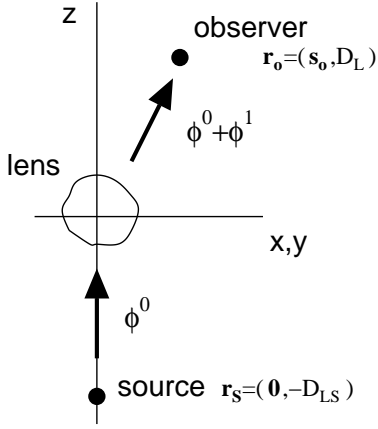}
  \caption{Gravitational lens geometry for the source, the lens and the
    observer. The lens is distributed around the origin.
    The source position is $\mathbf{r}_\mathrm{S}
    =(\mathbf{0},-D_\mathrm{LS})$, while
    the observer position is $\mathbf{r}_\mathrm{o}=(\mathbf{s}_\mathrm{o},
    D_\mathrm{L})$ where
    $\mathbf{s}_\mathrm{o}$ is a two-dimensional vector with
    $|\mathbf{s}_\mathrm{o}| \ll D_\mathrm{L}$.
    $D_\mathrm{L}$ and $D_\mathrm{LS}$ are the distances from the lens to
    the observer and to the source, respectively.
    $\phi^0$ is the incident wave, while $\phi^0 + \phi^1$ is
    the scattered wave.
}
  \label{fig1}
\end{figure}

We show the lens geometry of the source, the lens and the observer
 in Fig.\ref{fig1}.
The lens is distributed around the origin of the coordinate axes.
The source position is $\mathbf{r}_\mathrm{S}=(\mathbf{0},-D_\mathrm{LS})$,
 while the observer position is $\mathbf{r}_\mathrm{o}=(\mathbf{s}_\mathrm{o},
 D_\mathrm{L})$ where $\mathbf{s}_\mathrm{o}$ is a two-dimensional vector
 with $|\mathbf{s}_\mathrm{o}| \ll D_\mathrm{L}$.
 $D_\mathrm{L}$ and $D_\mathrm{LS}$ are the distances from the lens to the
 observer and to the source, respectively.

In the unlensed case $U=0$ in Eq.(\ref{weq}), we write
 $\tilde{\phi}^0(\mathbf{r})$ as the solution of this equation.
Including the effect of $U$ to first order, the
 scattered wave at the observer is written as
\beq
  \tilde{\phi}(\mathbf{r}_\mathrm{o}) = \tilde{\phi}^0 (\mathbf{r}_\mathrm{o})
    + \tilde{\phi}^1 (\mathbf{r}_\mathrm{o}),
\label{scw}
\eeq
where $\tilde{\phi}^1$ represents the effect of scattering.
Using Green's function for the Helmholtz equation,
 $\me^{\mi \omega \left| \mathbf{r} -\mathbf{r}^\prime \right|}
 /\left| \mathbf{r}-\mathbf{r}^\prime \right|$,
 we obtain $\tilde{\phi}^1(\mathbf{r}_\mathrm{o})$ from Eqs.(\ref{weq}) and
 (\ref{scw}) as
\beq
 \tilde{\phi}^1 (\mathbf{r}_\mathrm{o}) = - \frac{\omega^2}{\pi} \int d^3
 r^{\prime} \frac{\me^{\mi \omega \left| \mathbf{r}_\mathrm{o}
 -\mathbf{r}^\prime \right|}}
 {\left| \mathbf{r}_\mathrm{o} -\mathbf{r}^\prime \right|}
 U(\mathbf{r}^{\prime}) \tilde{\phi}^0 (\mathbf{r}^\prime).
\label{phi1}
\eeq
Note that the above result (\ref{phi1}) can be used if the lenses are broadly
 distributed between the source and the observer since the thin lens
 approximation is not assumed. 

We use the spherical wave emitted by the source as $\phi^0$,
 then we have $\tilde{\phi}^0(\mathbf{r}) = A \me^{\mi \omega \left|
 \mathbf{r}-\mathbf{r}_\mathrm{S} \right|}/\left| \mathbf{r}-
 \mathbf{r}_\mathrm{S} \right|$.  
In Fig.\ref{fig1}, $\phi^0$ represents the incident wave emitted by
 the source, while $\phi^0 + \phi^1$ represents the scattered wave.

\subsection{Geometrically thin lens}

We assume that the lens objects are locally distributed at the origin.
Then we have $\left| \mathbf{r}^\prime \right| \ll D_\mathrm{L,LS}$ and
 $\left| \mathbf{s}_\mathrm{o} \right| \ll D_\mathrm{L}$ in Eq.(\ref{phi1}).
Using the second-order Taylor series\footnote{$\left| \mathbf{r}-
 \mathbf{r}^\prime \right| \simeq r - \left( \mathbf{r} \cdot
 \mathbf{r}^\prime \right)/r + \left[{r^\prime}^2/r- \left(\mathbf{r} \cdot
 \mathbf{r}^\prime \right)^2/r^3 \right]/2$
 for $r \gg r^\prime$.} for these small quantities
 $\mathbf{r}^\prime$ and $\mathbf{s}_\mathrm{o}$,
 $\tilde{\phi}^1$ is reduced to
\beq
  \frac{\tilde{\phi}^1(\mathbf{r}_\mathrm{o})}
  {\tilde{\phi}^0(\mathbf{r}_\mathrm{o})}
 = -\frac{\omega^2}{\pi} \frac{D_\mathrm{S}}{D_\mathrm{L} D_\mathrm{LS}}
 \int d^3 r^\prime U(\mathbf{s}^\prime, z^\prime)
 \me^{\mi \omega t_\mathrm{d}(\mathbf{s}^{\prime},\mathbf{s}_\mathrm{o})},
\label{phi1-2}
\eeq
where $D_\mathrm{S}=D_\mathrm{L}+D_\mathrm{LS}$ and we set
 $\mathbf{r}^\prime=(\mathbf{s}^\prime, z^\prime)$.
Here, $t_\mathrm{d}$ is the geometrical time delay which is given by
 $t_\mathrm{d}(\mathbf{s}^{\prime},\mathbf{s}_\mathrm{o}) = {D_\mathrm{S}}
 /{\left( 2 D_\mathrm{L} D_\mathrm{LS} \right)} \times \left[
 \mathbf{s}^\prime - \left({D_\mathrm{LS}}/{D_\mathrm{S}}\right)
 \mathbf{s}_\mathrm{o} \right]^2$.
Using the two-dimensional gravitational potential $\psi(\mathbf{s}^\prime)
 = 2 \int dz^\prime U(\mathbf{s}^\prime, z^\prime)$, the result in
 Eq.(\ref{phi1-2}) is reduced to
\beq
  \frac{\tilde{\phi}^1(\mathbf{r}_\mathrm{o})}{\tilde{\phi}^0
 (\mathbf{r}_\mathrm{o})} = -\frac{\omega^2}{2 \pi} \frac{D_\mathrm{S}}
 {D_\mathrm{L} D_\mathrm{LS}} \int d^2 s^\prime \psi(\mathbf{s}^\prime)
 \me^{\mi \omega t_\mathrm{d}(\mathbf{s}^{\prime},\mathbf{s}_\mathrm{o})}.
\label{phi1-3}
\eeq
The above equation is also derived by expanding $\psi$ to first order
 in the diffraction integral (see Schneider \etal \cite{sef92}).
The surface density of the lens is defined as
 $\Sigma(\mathbf{s})=\int dz \rho (\mathbf{s},z)$, where $\rho$ is the
 mass density.  
Since the potential $\psi$ is written using the surface density as
 $\psi(\mathbf{s}^\prime)=4 \int d^2 s^{\prime \prime}
 \Sigma(\mathbf{s}^{\prime \prime}) \ln \left|\mathbf{s}^\prime
-\mathbf{s}^{\prime \prime}\right|$,
 the result (\ref{phi1-3}) is rewritten as (Appendix A),
\beqa
 \frac{\tilde{\phi}^1(\mathbf{r}_\mathrm{o})}{\tilde{\phi}^0
 (\mathbf{r}_\mathrm{o})}
 = 2 \mi \omega \int d^2 s^\prime \Sigma(\mathbf{s}^\prime) \left[ E_\mathrm{i}
 \left( \mi \omega t_\mathrm{d}(\mathbf{s}^{\prime},\mathbf{s}_\mathrm{o})
 \right) - \ln \left| \mathbf{s}^\prime - \frac{D_\mathrm{LS}}{D_\mathrm{S}}
 \mathbf{s}_\mathrm{o} \right|^2 \right], 
\label{phi1-4}  \nonumber \\
\eeqa
where $E_\mathrm{i}$ is the exponential integral function : $E_\mathrm{i}
 (\mi x)=- \int_x^\infty dt~\me^{\mi t}/t$ (e.g. Abramowitz and Stegun
 \cite{as70}).
If we shift the potential $\psi$ by a constant value $\psi_0$, 
 $\tilde{\phi}^1 / \tilde{\phi}^0$ in Eq.(\ref{phi1-3}) is shifted by
 $-\mi \omega \psi_0$.\footnote{In other words, this corresponds to
 an additional phase shift in the incident wave : $\tilde{\phi}^0
 \to \tilde{\phi}^0 \me^{-\mi \omega \psi_0} \simeq \tilde{\phi}^0
 \left( 1-\mi \omega \psi_0 \right)$.
 But, $\tilde{\phi}^1$ is not changed under this shift since
 $\tilde{\phi}^1 \psi_0 \approx \mathcal{O}(U^2)$ is negligible.
 Hence, we are free to choose $\psi_0$.}
Hence we can choose $\psi_0$ so that the second term of
 Eq.(\ref{phi1-4}) vanishes, and we eliminate this term.

Let us discuss the validity of the Born approximation.
If the scattered wave $\tilde{\phi}^0+\tilde{\phi}^1$ is not too
 different from the incident wave $\tilde{\phi}^0$, this approximation
 is valid.
We discuss the condition of $|\tilde{\phi}^1/\tilde{\phi}^0| \ll 1$
 for the two lens models, point mass lens and smoothly distributed lens,
 in the following subsections.

\subsubsection{Point mass lens}

\begin{figure}
  \includegraphics[width=7.cm]{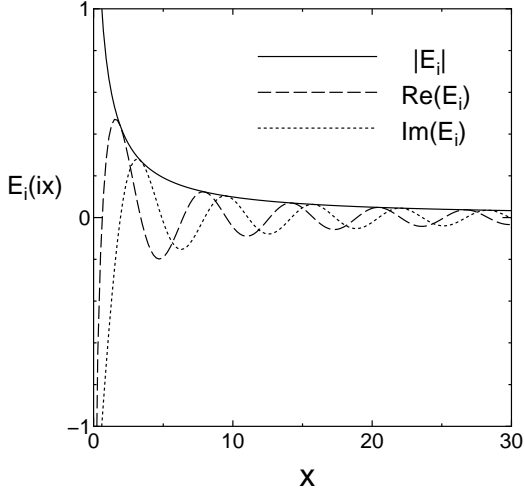}
  \caption{$E_\mi(\mi x)$ as a function of $x$. The solid line is the
 absolute value of $E_\mi$, the dashed line is the real part, and the
 dotted line is the imaginary part.
}
  \label{fig2}
\end{figure}

The surface density is $\Sigma(\mathbf{s})=M \delta^2(\mathbf{s})$
 where $M$ is the lens mass.
Then, $\tilde{\phi}^1$ in Eq.(\ref{phi1-4}) is rewritten as
 (see also Peters \cite{p74}, Sect. III A)
\beq
   \frac{\tilde{\phi}^1(\mathbf{r}_\mathrm{o})}{\tilde{\phi}^0
 (\mathbf{r}_\mathrm{o})} = 2 \mi M \omega ~E_\mathrm{i}
  \left( \frac{\mi \omega D_\mathrm{LS}}{2 D_\mathrm{L} D_\mathrm{S}}
 s_\mathrm{o}^2 \right).
\label{point}
\eeq 
In Fig.\ref{fig2}, we show $E_\mathrm{i}(\mi x)$ as a function of $x$. 
The solid line is the absolute value of $E_\mathrm{i}$, the dashed line is
 the real part, and the dotted line is the imaginary part.
From this figure, except for small $x \lsim 1$, $\left| E_\mathrm{i} \right|$
 is smaller than $1$. Thus the Born approximation is valid 
 $\left| \tilde{\phi}^1 / \tilde{\phi}^0 \right| \ll 1$
 in Eq.(\ref{point})
 if the lens mass $M$ is smaller than the wave length of the
 gravitational waves $\lambda = 2 \pi/\omega$ except for small
 impact parameter $s_\mathrm{o} \lsim \left( D_\mathrm{L} D_\mathrm{S}/\omega
 D_\mathrm{LS} \right)^{1/2}$.

Especially for large impact parameter $s_o$,
 since $E_\mi(\mi x) \sim -\mi \me^{\mi x}/x$ for 
 $x \gg 1$, Eq.(\ref{point}) is reduced to 
\beq
   \frac{\tilde{\phi}^1(\mathbf{r}_\mo)}{\tilde{\phi}^0(\mathbf{r}_\mo)}
 = \left( \frac{s_\mE}{s_\mo} \right)^2
 \exp \left( \frac{\mi \omega D_\mLS}{2 D_\mL D_\mS} s_\mo^2 \right),
\label{point2}
\eeq 
where $s_\mE$ is the Einstein radius on the observer plane :
 $s_\mE = \left( 4 M D_\mL D_\mS /D_{\mLS} \right)^{1/2}$.
Hence if the observer position $s_\mo$ is larger than the Einstein
 radius $s_\mE$, the Born approximation is valid irrespective of the
 wavelength of the gravitational waves.

\subsubsection{Smoothly distributed lens}

We assume that the lenses are  smoothly distributed on the $z=0$ plane
 in Fig.\ref{fig1}.
In the unlensed case, the wave propagates through 
 $\mathbf{s}^\prime = (D_{\mLS}/D_\mS) \mathbf{s}_\mo$ in the lens plane.
Expanding the lens potential $\psi(\mathbf{s}^\prime)$ around it,
 $\tilde{\phi}^1$ in Eq.(\ref{phi1-3}) is rewritten as
\beq
   \frac{\tilde{\phi}^1(\mathbf{r}_\mo)}{\tilde{\phi}^0(\mathbf{r}_\mo)}
 = -\mi \omega \psi + \kappa
 + \frac{\mi}{4 \omega} \frac{D_\mL D_{\mLS}}{D_\mS} \nabla^2_\ms \kappa
 + \mathcal{O} \left[ \left( \frac{D_\mL D_{\mLS}}{\omega D_\mS} \right)^2
 \nabla^4_\ms \kappa \right].
\label{weakphi}
\eeq 
The first term, being $-\mi \omega \psi(\mathbf{s}^\prime)$ at
 $\mathbf{s}^\prime = (D_{\mLS}/D_\mS) \mathbf{s}_\mo$, can be eliminated for
 the same reason as the second term in Eq.(\ref{phi1-4}).   
In the second term, $\kappa$ is the convergence field along the line of
 sight to the source defined as $\kappa=\Sigma(\mathbf{s}^\prime)
 /\Sigma_\mathrm{cr}$ at $\mathbf{s}^\prime = ( D_{\mLS}/D_\mS )
 \mathbf{s}_\mo$ with $\Sigma_\mathrm{cr}=D_\mS/\left( 4 \pi D_\mL D_{\mLS}
 \right)$.
The third term is a correction term arising from the effect of finite
 wavelength.
If the scale of the density fluctuation $\sim
 \left| \nabla^2_\ms \kappa/\kappa \right|^{-1/2}$ is much larger than
 the Fresnel scale $\sim \left( D_\mL D_{\mLS}/\omega D_\mS \right)^{1/2}$,
 or if the geometrical optics is valid $(\omega \to \infty)$,
 the third term is negligible.
Then, the incident wave is magnified by $1+\kappa$, which is consistent
 with the result in the weak gravitational lensing
 (e.g. Bartelmann \& Schneider \cite{bs01}).      
If $\kappa \ll 1$, the Born approximation is valid.

\subsection{Geometrically thick lens}

We consider the lenses are broadly distributed between the source and
 the observer.
It is easy to apply the previous result in Sect. 2.1 to this case.
$\tilde{\phi}^1$ in Eq.(\ref{phi1-3}) is rewritten as
\beqa
  \frac{\tilde{\phi}^1(\mathbf{r}_\mo)}{\tilde{\phi}^0(\mathbf{r}_\mo)}
 &=& -\frac{\omega^2}{\pi} \int d^2 s^\prime \int_{0}^{D_\mS}
 d z^\prime \frac{D_\mS}{\left( D_\mS-z^\prime \right) z^\prime} \nonumber \\
 && \hspace*{1.5cm} \times  U(\mathbf{s}^\prime,
 z^\prime-D_{\mLS}) ~\me^{\mi \omega t_\mathrm{d}(\mathbf{s}^{\prime},
 z^\prime, \mathbf{s}_\mo)},
\label{brd-phi}
\eeqa
with
  $t_\mathrm{d}(\mathbf{s}^{\prime}, z^\prime, \mathbf{s}_\mo) =
 D_\mS/ \left[ 2 z^\prime \left( D_\mS-z^\prime \right) \right] 
 \times \left[ \mathbf{s}^\prime -
 ({z^\prime}/{D_\mS}) \mathbf{s}_\mo \right]^2$.
If the lenses have a thickness of $\Delta z$ in the $z$ direction, 
 the correction for this thickness in the scattered wave
 $\tilde{\phi}^1$ is of the order of $\Delta z/D_\mS$ from
 Eqs. (\ref{phi1-2}) and (\ref{brd-phi}).
Thus, if $\Delta z \ll D_\mS$ the thin lens approximation is valid.   

Especially, for smoothly distributed lenses the result (\ref{weakphi}) is
 valid but $\kappa$ is replaced by
\beq
\kappa = 4 \pi \int_0^{D_\mS} d z^\prime \frac{z^\prime \left( D_\mS-z^\prime
 \right)} {D_\mS} ~\rho \left( ({z^\prime}/{D_\mS}) \mathbf{s}_\mo,
  z^\prime - D_{\mLS} \right),
\eeq
where $\rho$ is the mass density of the lens.

\subsection{Two-point correlation function}

Recently, Macquart (\cite{m04}) (hereafter M04) derived the correlation
 function in the wave amplitude of the two detectors under the thin lens
 approximation.
He suggested that measurement of the correlation function provides
 the power spectrum of the mass density fluctuation. 
In this section, we derive it without the thin lens approximation,
 but within the limit of weak gravitational fields.\footnote{
The correlation function due to electromagnetic scattering was
 exactly obtained, not only for weak fluctuation but also for
 strong fluctuation. See references, Ishimaru \cite{i78},
 Tatarskii \& Zavorotnyi \cite{tz80}, for a detailed  discussion.} 
We consider two observers at $\mathbf{r}_\mo$ and $\mathbf{r}_\mo
 + \mathbf{r}_1$ with $\left| \mathbf{r}_1 \right| \ll \left| \mathbf{r}_\mo
 \right|$.
The mass fluctuation is usually characterized by the power spectrum
 defined as $P(k)=\int \langle \rho (\mathbf{r}) \rho (\mathbf{r} +
 \mathbf{r}^\prime) \rangle \me^{\mi \mathbf{k} \cdot \mathbf{r}^\prime}
 d^3 r^\prime$.
The correlation in the potential $U(\mathbf{r})$ and
 $U(\mathbf{r}^\prime)$ is written as
\beq
  \langle U(\mathbf{r}) U(\mathbf{r}^\prime) \rangle = \frac{2}{\pi}
 \int d^3 k ~\frac{1}{k^4} ~P(k) ~\me^{-\mi \mathbf{k} \cdot \left(
 \mathbf{r} - \mathbf{r}^\prime \right)}.
\label{coru}
\eeq 
We obtain the correlation in the wave amplitudes
 $\tilde{\phi}^1(\mathbf{r}_\mo)$ and $\tilde{\phi}^1(\mathbf{r}_\mo
 + \mathbf{r}_1)$ from Eqs.(\ref{phi1}) and (\ref{coru}) as
\beqa
 && \langle \tilde{\phi}^1(\mathbf{r}_\mo) \tilde{\phi}^{1 *} (\mathbf{r}_\mo+
 \mathbf{r}_1) \rangle = \frac{2 \omega^4}{\pi^3} \int d^3 k ~\frac{1}{k^4}
 ~P(k) \int d^3 r^\prime \int d^3 r^{\prime \prime}  \nonumber \\
 &&  ~~~~\times \frac{\me^{\mi \omega \left| \mathbf{r}_\mo -\mathbf{r}^\prime
 \right|}}{\left| \mathbf{r}_\mo-\mathbf{r}^\prime \right|}
 ~\tilde{\phi}^0(\mathbf{r}^\prime)
 ~\frac{\me^{- \mi \omega \left| \mathbf{r}_\mo + \mathbf{r}_1
 -\mathbf{r}^{\prime \prime} \right|}}{\left| \mathbf{r}_\mo + \mathbf{r}_1
 -\mathbf{r}^{\prime \prime} \right|}
 ~\tilde{\phi}^{0 *}(\mathbf{r}^{\prime \prime}) ~\me^{-\mi \mathbf{k}
 \cdot \left( \mathbf{r}^\prime - \mathbf{r}^{\prime \prime} \right)}.
\label{cor-phi}
\eeqa
It is useful to change the integral variables to $\mathbf{x}=\mathbf{r}^\prime
 -\mathbf{r}^{\prime \prime}$ and $\mathbf{y}=\left( \mathbf{r}^\prime
 +\mathbf{r}^{\prime \prime} \right)/2$, and we assume that $\mathbf{x}$ is
 much smaller than the distances $D_{L,LS}$.
Then, expanding these small quantities $\mathbf{x}$ and $\mathbf{r}_1$ in
 the exponentials, Eq.(\ref{cor-phi}) is rewritten as
\beqa
 && \langle \tilde{\phi}^1(\mathbf{r}_\mo) \tilde{\phi}^{1 *} (\mathbf{r}_\mo+
 \mathbf{r}_1) \rangle = \frac{2 \omega^4 A^2}{\pi^3} \int d^3 k ~\frac{1}{k^4}
 ~P(k) \int d^3 x \int d^3 y  \nonumber \\
 &&  ~~~~\times \frac{1}{\left| \mathbf{r}_\mo - \mathbf{y} \right|^2
 \left| \mathbf{r}_\mS - \mathbf{y} \right|^2}
 ~\me^{-\mi \omega \left[ \left( \widehat{\mathbf{r}_\mo
 - \mathbf{y}} + \widehat{\mathbf{r}_\mS - \mathbf{y}} \right) \cdot
 \mathbf{x} + \left( \widehat{\mathbf{r}_\mo - \mathbf{y}} \right) \cdot
 \mathbf{r}_1 \right]} ~\me^{-\mi \mathbf{k} \cdot \mathbf{x}}, 
\label{cor-phi1}
\eeqa
where $\widehat{\mathbf{v}}$ denotes the unit vector for $\mathbf{v}$.
$A$ is the amplitude of the incident wave (see sentences after
 Eq.(\ref{phi1})).
Performing the integral in  Eq.(\ref{cor-phi1}), we obtain
\beqa
  \langle \tilde{\phi}^1(\mathbf{r}_\mo) \tilde{\phi}^{1 *} (\mathbf{r}_\mo+
 \mathbf{r}_1) \rangle = 16 \omega^2 \left| \tilde{\phi}^0(\mathbf{r}_\mo)
 \right|^2 \int d^2q \frac{1}{q^4} ~P(q) \nonumber \\
 \times \int_0^{D_\mS} dz^\prime
 \me^{- \mi \omega \mathbf{n} \cdot \mathbf{r}_1},
\label{cor-phi2}
\eeqa
where $\mathbf{n} = \left( - z^\prime \mathbf{q} / \omega D_S, 1 \right)$ and
 $\mathbf{q}$ is the $(x,y)$ component of the wave vector $\mathbf{k}$.
In the geometrically thin lens at $z=0$ plane, $P(q)$ is
 replaced by $P^{2D}(q) \delta(z^\prime-D_{\mLS})$ where $P^\mathrm{2D}$ is
 $\mathrm{2D}$ power spectrum.
  
Let us consider a single power law model for the power spectrum
 as an example.
This model can be used for the fluctuation of cold dark matter
 and gas (M04, Sec.5.2 \& 5.3).
The power spectrum is $P(k)=P_0 \left( k/k_0 \right)^{-\mn}$
 for $k_\mathrm{min}<k<k_\mathrm{max}$, and $P(k)=0$ otherwise.
The index is $\mn \approx 3-4$, and $k_\mathrm{min} \ll k_\mathrm{max}$. 
Then, the integral in Eq.(\ref{cor-phi2}) is dominated by the fluctuation
 at the largest scale of $1/k_\mathrm{min}$.
The exact solution of integral (\ref{cor-phi2}) was given by the
 hypergeometric function, but we present an approximate solution
 for simplicity.
Assuming that the separation of two detectors $|\mathbf{r}_1|$ is much
 smaller than the largest scale of fluctuation $1/k_\mathrm{min}$, we have
\beqa
  \langle \tilde{\phi}^1(\mathbf{r}_\mo) \tilde{\phi}^{1 *} (\mathbf{r}_\mo+
 \mathbf{r}_1) \rangle = 16 \omega^2 \left| \tilde{\phi}^0(\mathbf{r}_\mo)
 \right|^2 P(k_\mathrm{min}) k_\mathrm{min}^{-2} D_\mS \me^{-\mi \omega z_1}
  \nonumber \\
   \times \left[ \frac{2\pi}{2-\mn} + \frac{\pi}{6 \mn}
 \left( k_\mathrm{min} s_1 \right)^2 +
 \mathcal{O} \left(k_\mathrm{min}^4 s_1^4\right) \right],
\label{cor-phi3}
\eeqa
where $\mathbf{r}_1=(\mathbf{s}_1,z_1)$.
The first term represents the dispersion of the scattered wave,
 while the second term is two-point correlation function.\footnote{
 We comment on the first term. In the previous work (M04),
 he includes the correlation between the scattered wave being second order
 of potential $\mathcal{O}(\psi^2)$ and the incident wave.  
Then, the correlation at $s_1=0$ is subtracted (see Eq.(23) in his
 paper), and $\me^{-\mi \omega \mathbf{n} \cdot \mathbf{r}_1}$ in
 Eq.(\ref{cor-phi1}) is replaced by $\me^{-\mi \omega \mathbf{n} \cdot
 \mathbf{r}_1} - \me^{-\mi \omega z_1}$.
Hence, the first term would vanish if we included the second order of
 $\psi$ in the scattered wave.}
If the lenses have a finite thickness $\Delta L$ along the z-axis,
 $D_S$ in Eq.(\ref{cor-phi3}) should be replaced by $\Delta L$.

\section{Conclusions}

We have discussed the scattering of gravitational waves by the weak
 gravitational fields of lenses by using the Born approximation.
We consider the two lens models, the point mass lens and the
 smoothly distributed lens, and discuss the validity of the Born
 approximation.
For the point mass lens, the effect of scattering is roughly given by
 the Schwarzschild radius $M$ of lens divided by the wavelength $\lambda$. 
If $M < \lambda$ or if the impact parameter
 is larger than the Einstein radius, the approximation is valid.
For the smoothly distributed lens, the effect of scattering is of the
 order of the convergence $\kappa$, and if $\kappa \ll 1$ the approximation
 is valid.
In the short wavelength limit, the result is consistent with the
 weak gravitational lensing.   
We derive the correction term due to the effect of finite wavelength in 
 the magnification.
The two point correlation function is also discussed following the recent
 paper (M04).

\begin{acknowledgements}
We would like to thank the referee for useful comments to improve
 the manuscript. 

\end{acknowledgements}

\appendix

\section{Derivation of Eq.(\ref{phi1-4})}

Inserting $\psi(\mathbf{s}^\prime)=4 \int d^2 s^{\prime \prime}
 \Sigma(\mathbf{s}^{\prime \prime}) \ln \left| \mathbf{s}^\prime
-\mathbf{s}^{\prime \prime} \right|$ into Eq.(\ref{phi1-3}), we change the
 integral variable from $\mathbf{s}^\prime$ to $\mathbf{u} \equiv
 \mathbf{s}^\prime - \mathbf{s}^{\prime \prime}$.
Then, we have
\beqa
 \frac{\tilde{\phi}^1(\mathbf{r}_\mo)}{\tilde{\phi}^0(\mathbf{r}_\mo)}
 = -4 \omega^2 \frac{D_\mS}{D_\mL D_{\mLS}} \int d^2 s^{\prime \prime}
 \Sigma(\mathbf{s}^{\prime \prime}) \me^{\mi \omega t_\mathrm{d}
 (\mathbf{s}^{\prime \prime}, \mathbf{s}_\mo)} \nonumber \\
 \times \int_0^\infty du ~u \ln u ~\me^{\mi \alpha u^2} J_0(\beta u),
\label{app1}
\eeqa
where $\alpha=\omega D_\mS / \left( 2 D_\mL D_{\mLS} \right)$, $\beta=2 \alpha
 \left| \mathbf{s}^{\prime \prime}-(D_{\mLS}/D_\mS) \mathbf{s}_\mo \right|$,
 and $J_0$ is the 0-th order Bessel function.
By using series $J_0 (\beta u)=\sum_{\mn=0}^{\infty}
 \left(- \beta^2 u^2/4 \right)^\mn \left( \mn! \right)^{-2}$, the integral
 in Eq.(\ref{app1}) is rewritten as,
\beqa
 \int_0^\infty du ~u \ln u ~\me^{\mi \alpha u^2} J_0(\beta u) = -
 \frac{\mi}{4 \alpha} \left[ \gamma + \ln \left(- \mi \alpha \right)
 \right] \nonumber \\
 + \frac{\mi}{4 \alpha} \sum_{\mn=1}^\infty \frac{1}{\mn!} \left( -
 \frac{\mi \beta^2}{4 \alpha} \right)^\mn  \left[ \sum_{\mk=1}^\mn
 \frac{1}{\mk} - \gamma - \ln \left( -\mi \alpha \right) \right],
\label{app2}
\eeqa
where $\gamma$ is the Euler constant (e.g. Gradshteyn \& Ryzhik \cite{gr00}).
Using identity $\sum_{\mk=1}^\mn 1/\mk=\int_0^1 dt [1-(1-t)^\mn]/t$, we have
\beqa
 \sum_{\mn=1}^\infty \frac{1}{\mn!} \left( - \frac{\mi \beta^2}{4 \alpha}
 \right)^\mn \sum_{\mk=1}^\mn \frac{1}{\mk} =\me^{-\mi \beta^2/(4 \alpha)}
 \int_0^1 \frac{dt}{t} \left( 1- \me^{ \mi \beta^2 t /(4 \alpha)} \right)
 \nonumber \\
 =  -\me^{-\mi \beta^2/(4 \alpha)} \left[ E_\mi \left(\mi \beta^2/(4 \alpha)
 \right) - \ln \left( -\mi \frac{\beta^2}{4 \alpha} \right) -\gamma \right].
\label{app3}
\eeqa
In the second equality, we use $E_\mi(\mi x)=\gamma+\ln \left( -\mi x \right) 
 +\sum_{\mn=1}^\infty \left( \mi x \right)^\mn / \left( \mn \cdot \mn!
 \right)$.
Inserting Eqs.(\ref{app2}) and (\ref{app3}) into Eq.(\ref{app1}), we obtain
 Eq.(\ref{phi1-4}).

\end{document}